\documentclass[dvips]{article}
\usepackage{icrctc07}

\title{A First Synoptic Blazar Study Comprising Thirteen Blazars Visible in
$E>100$ GeV Gamma-Rays}
\shorttitle{A Synoptic VHE Blazar Study}

\authors{Robert Wagner$^{1}$ }
\shortauthors{Wagner et al.}
\afiliations{$^1$Max-Planck-Institut f\"{u}r Physik, F\"{o}hringer
Ring 6, D--80805 M\"{u}nchen, Germany }
\email{robert.wagner@mppmu.mpg.de}

\abstract{Since 2002, the number of detected blazars at $E_\gamma
> 100$~GeV has more than doubled. I study all currently known
BL Lac-type objects with published energy
spectra. Their intrinsic energy spectra are reconstructed by
removing extragalactic background light attenuation effects. The
emission properties are then compared and correlated among each
other, with X-ray data, and with the individual black hole masses.
In addition, I consider temporal properties of the very high
energy $\gamma$-ray flux. Key findings concern the flux--black
hole mass and variability scale--black hole mass connections and
the correlation of the spectral slope and the luminosity.  As a
specific application, the study allows to constrain the still
undetermined redshift of the blazar PG 1553+113.}

\begin{document}
\maketitle
\section{Introduction and Approach}
In order to assess both the acceleration mechanisms in blazars and
extragalactic background light (EBL) absorption effects, not only
individual object studies, but also the investigation of a large
sample of very high energy (VHE) $\gamma$-ray emitting blazars is
desirable. Ideally it should encompass a wide range in redshift
for EBL studies and also include groups of sources at comparable
distances in order to study intrinsic properties of the individual
sources without possible systematic uncertainties caused by the
EBL de-absorption.
The preconditions for such comparative blazar studies have much
improved recently. To date, the VHE $\gamma$-ray blazar sample
comprises 18 BL Lac objects (e.g. \cite{up95}),
with $z= 0.030\dots0.212$.\footnote{See
http://www.mppmu.mpg.de/$\sim$rwagner/sources/ for an up-to-date list.}
A sample of all blazars with reported energy spectra is studied by inferring
the intrinsic emission properties of the individual objects and by
probing correlations of their VHE $\gamma$-ray and X-ray emission
properties with black hole (BH) mass ($M_\bullet$) estimates.

\paragraph{Black Hole Masses.}

We estimate $M_\bullet$ by evaluating the $M_\bullet$-$\sigma$
relation \cite{2002ApJ...574..740T}, i.e. the tight correlation of
the stellar velocity dispersion $\sigma$ and $M_\bullet$ of nearby
galaxies. This approach assumes that AGN host galaxies are similar
to inactive galaxies as far as the $M_\bullet$-$\sigma$ relation
is concerned. We find that the currently VHE $\gamma$-ray emitting BL~Lacs
are flatly distributed in
$M_\bullet=(10^8-10^{9.5}) \mathrm{M}_\odot$. Note that although
AGNs harbor BHs with $M_\bullet
> 10^6 \mathrm{M}_\odot$, up to now only BL~Lacs with $M_\bullet > 10^8 \mathrm{M}_\odot$
have been discovered in VHE $\gamma$-rays.

  \begin{figure*}[t]
    \hfill
    \begin{minipage}[t]{.3\textwidth}
      \begin{center}
        \includegraphics[width=.8\textwidth]{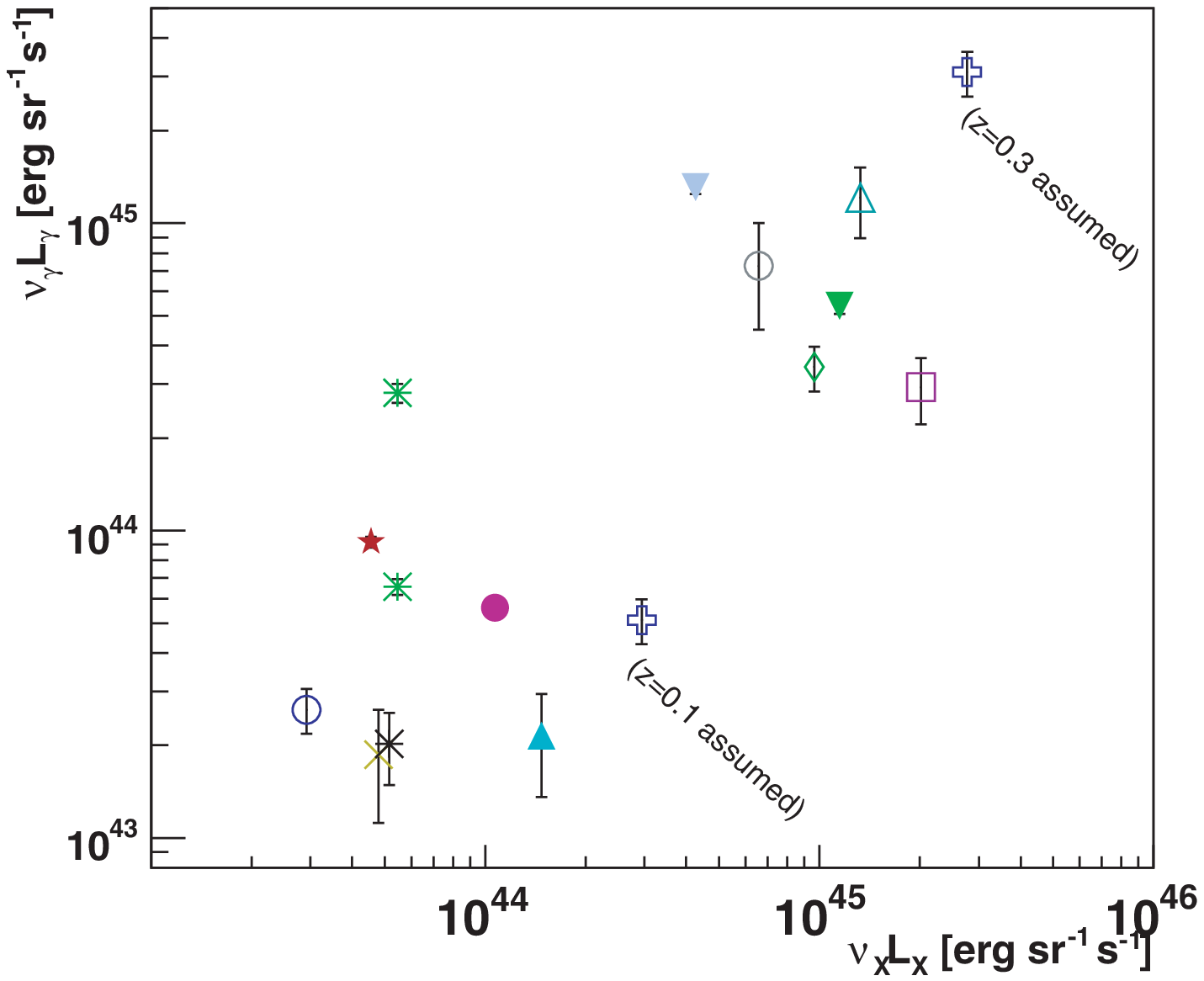}
        \caption{$\nu_\gamma L_\gamma$ vs. $\nu_X L_X$. The symbols are explained in Fig.~2. The PG\,1553 points are for assumed $z=0.1$ and $z=0.3$.}
        \label{fig:comp:compaxrc}
      \end{center}
    \end{minipage}
    \hfill
    \begin{minipage}[t]{.6\textwidth}
      \begin{center}
        \includegraphics[width=.9\textwidth]{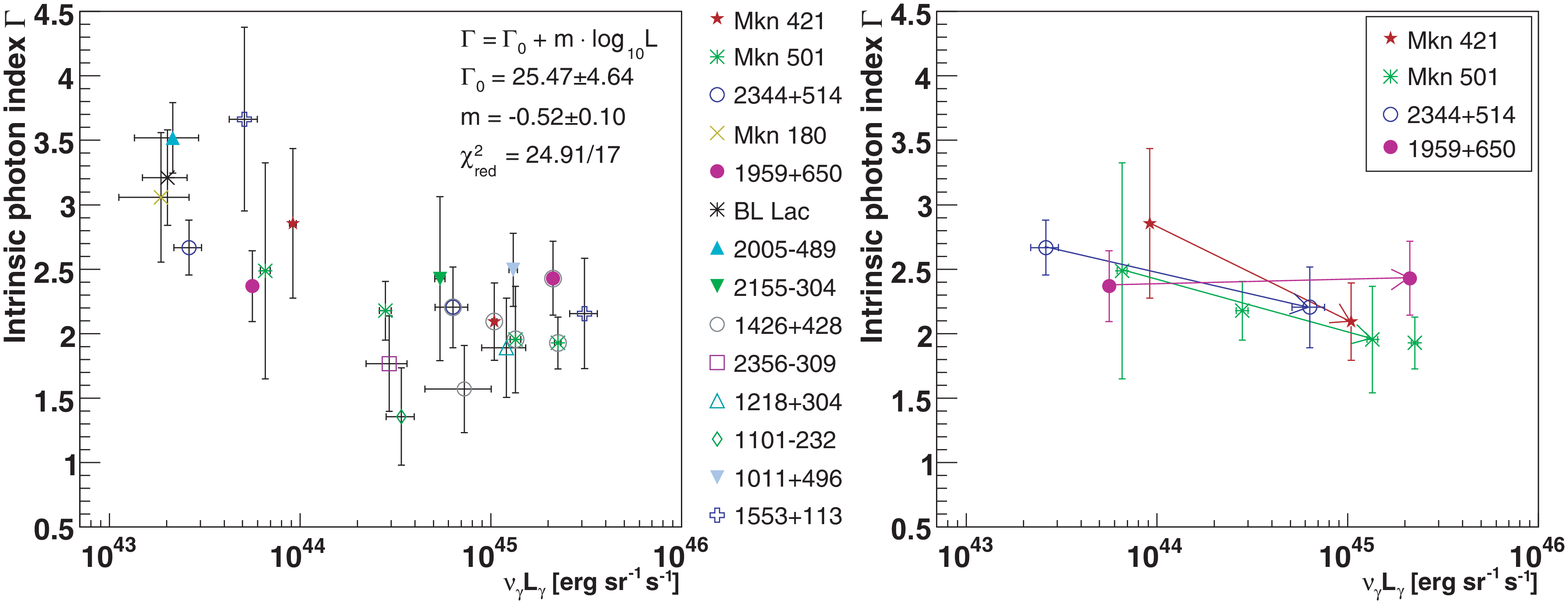}
        \caption{Photon index vs. $\nu_\gamma L_\gamma$. Additional flare states of sources are marked by gray circles.
        The PG\,1553 points were not included in the fit. Right: Four blazars with low and high states.}
        \label{fig:comp:compasllum}
      \end{center}
    \end{minipage}
    \hfill
  \end{figure*}

\paragraph{Intrinsic VHE $\gamma$-Ray Emission Parameters.}

The photon spectra measured in the VHE range suffer absorption by
EBL \cite{HauserDwek}. Here, the intrinsic blazar spectra are
reconstructed using the EBL ``low'' model given in
\cite{kneiske4}. For sources that have been found in different
flux states, ``low state'' and ``flare'' spectra are considered.
Data from Mkn 421 \cite{421}, Mkn~501 \cite{501}, 1ES\,2344
\cite{2344}, 1ES\,1959 \cite{1959}, PKS~2155 \cite{Aharonian2005},
1H~1426 \cite{1426}, PKS~2005 \cite{HESS2005}, 1ES\,1218
\cite{MAGIC1218}, 1ES\,2356 \& 1ES\,1101 \cite{HESSAGNNature},
PG\,1553 \cite{magic1553}, Mkn~180 \cite{MAGIC180}, PKS\,0548
\cite{548}, BL Lac \cite{bllac}, and 1ES\,1011 \cite{1011} have been included.
Throughout this study, the unknown-redshift object PG\,1553+113 is
assumed at several possible $z$ values, but not included further
unless explicitly stated otherwise. The extracted observables are
the intrinsic luminosity at 500 GeV and the intrinsic photon index
$\Gamma$ in the region around 500~GeV. For both, no extrapolations
beyond the spectral fits are required. The resulting intrinsic
photon indices vary from $\Gamma=1.5 - 3.3$. In most current
acceleration models only $\Gamma > 1.5$ is allowed. This goes in
line with indications that the EBL absorption effects are still
smaller than modeled \cite{HESSAGNNature}.

\section{Correlation of X-ray and $\gamma$-Ray Luminosity}
In SSC models, the X-ray and the VHE emission are closely
connected,  owing to their common origin.
Fig.~\ref{fig:comp:compaxrc} shows $\nu_\gamma L_\gamma$ versus
the X-ray luminosity at 1 keV ($\nu_X L_X$; from Costamante \&
Ghisellini 2002). A trend towards a correlation is visible, even
though a strong correlation might not even be expected due to
different magnetic fields in the individual objects.
Note that high thermal contributions at 1 keV are unlikely and
would imply a very high amount of gas and pressure.

\section{Correlation between Photon Index and $\gamma$-ray Luminosity}
Fig.~\ref{fig:comp:compasllum} relates the intrinsic photon
indices $\Gamma$ to $\nu_\gamma L_\gamma$. A correlation on the $3.3\sigma$
level is found, which, within SSC models, is compatible with a
moving IC peak towards higher energies and an IC peak energy
$<500$\,GeV. Sources with observed spectra at individual distinct
flux states support this correlation. Mkn\,501 and 1ES\,2344 show
a similar change in spectral slope and a luminosity increase of
$\Delta(\nu_\gamma L_\gamma) \approx 20$. The luminosity increase of Mkn~421 is much
lower with $\Delta(\nu_\gamma L_\gamma) \approx 10$.

\section{Correlations of $\gamma$-Ray Emission with the BH Properties and $z$}
The properties of blazar $\gamma$-ray emission are expected to be
connected to BH properties, like $M_\bullet$ and its spin, since
scaling laws govern BH physics, in particular length and time
scales \cite{1999ARA&A..37..409M}. Currently, only $M_\bullet$ can
be reliably estimated. The BH spin remains inaccessible by large;
the accretion rate might be indirectly accessible through the
(radio) jet power.
A first study of the connection of source properties and
$M_\bullet$ of the then-established five TeV blazars
\cite{Krawczynski1es1959-2004} did not find any correlations,
except for an indication of a connection between the X-ray flare
duty cycle and $M_\bullet$ (see below).

\begin{figure*}
\center{\includegraphics[width=\linewidth]{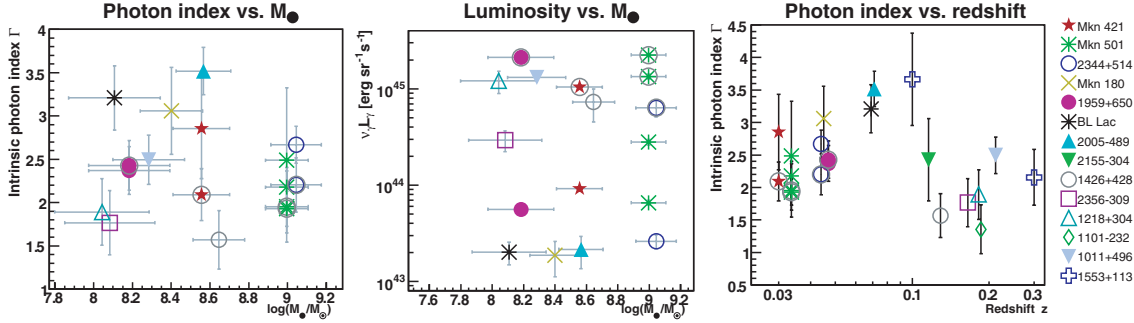}
\caption{Correlations of $\Gamma$, $\nu_\gamma L_\gamma$ with $M_\bullet$,
$z$. PG\,1553 at assumed $z=0.1$ and $z=0.3$.
\label{fig:comp:compaspecs2}}}
\end{figure*}

Fig.~\ref{fig:comp:compaspecs2} shows the correlation of $\Gamma$
and $\nu_\gamma L_\gamma$ with $M_\bullet$ and also tests for possible
correlations with $z$. The latter are not expected from physics,
but may identify selection effects in the data sample and/or an
inaccurate EBL model.
Only sources with hard intrinsic spectra are visible at large
distances ($z>0.1$), because soft spectra more easily fall below
the current instrumental sensitivity limits. Another explanation
for the prevalent hard spectra at large $z$ is an overcorrection
of the EBL attenuation effects. None of the nearby sources, for
which no strong EBL modifications apply, show $\Gamma$ much
smaller than 2.0. Additionally, the detected number of objects
with soft spectra increased substantially since 2002.

While there is no obvious correlation between $M_\bullet$ and the
VHE $\gamma$-ray luminosity, it might be that the current data
populate only a certain area in the $M_\bullet$---$\Gamma$ plane.
Owing to the large uncertainties of the $M_\bullet$ determination
and the still poor statistics, the future will have to show if
such trends are real. Perhaps the VHE $\gamma$-ray emission is
more sensitive to the BH spin, the accretion rate or, more
importantly, of the acceleration environment rather than the BH
mass. Also results on timing properties (see below) support such
claims.

\section{The Case of PG\,1553+113}

\begin{figure}[h!]
\center{\includegraphics[width=\linewidth]{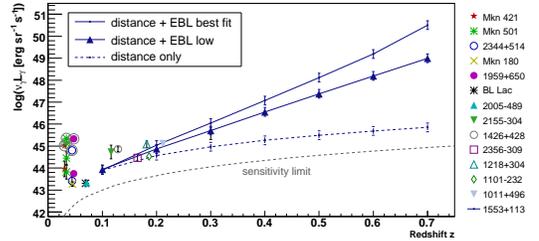}
\caption{Luminosity evolution for PG\,1553+113 at different
assumed source distances.\label{fig:comp:1553lumilim}}}
\vspace{-0.4cm}
\end{figure}

PG\,1553 is a recently discovered TeV blazar \cite{magic1553} with
unknown distance. With increasing redshift $z$, the intrinsic
luminosity has to increase stronger than quadratic
($\nu_\gamma L_\gamma \geq 4 \pi \cdot d^2 F$) due to EBL absorption as to sustain
the measured VHE flux (Fig.~\ref{fig:comp:1553lumilim}). We assume
here that PG\,1553+113 is an ``off the shelf'' blazar, i.e. with
no extraordinarily high $L$. This assumption is difficult to
quantify, but when translating it into the limit that $L$ is not
more than 30 times higher than the highest luminosities observed,
one obtains $z<0.48$ ($2\sigma$ limit). An extreme luminosity
1000-times higher yields a limit of $z<0.68$. Among the extreme
BL~Lac objects we find $(\nu_\gamma L_\gamma)_\mathrm{max}<10^{45.4}
~\mathrm{erg} ~\mathrm{sr}^{-1} ~\mathrm{s}^{-1}$ for Mkn~501 in
the flare state. These limits do not only depend on a good
knowledge of the EBL attenuation over a wide range in redshift,
but also on an assumed reasonable maximum VHE blazar luminosity
that is strongly dependent on the jet Doppler factor $\delta$. In
any case, either a strikingly high luminosity or a very high $\delta$ is
needed should PG\,1553+113 be more distant than $z>0.35$.
Presumably such very extreme objects are so rare that a
sufficiently large volume had to be probed to find one of them.

\section{X-ray Duty Cycle and VHE Variability Time Scale}

\begin{figure}[h!]
\center{\includegraphics[width=\linewidth]{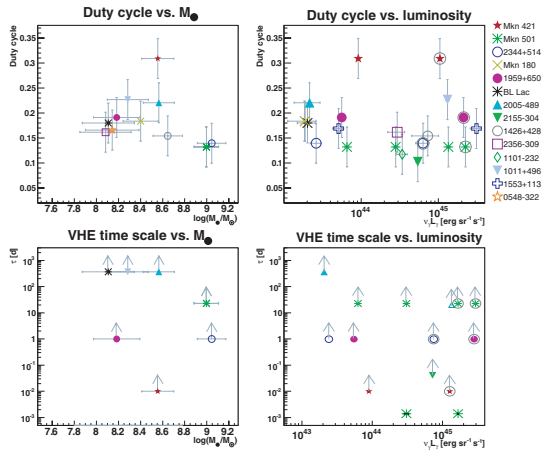}
\caption{Correlations of X-ray duty cycle and VHE variability
scales with $M_\bullet$ and $\nu_\gamma L_\gamma$.
\label{fig:comp:compaspecstime}}} \vspace{-0.5cm}
\end{figure}

Following a method described in \cite{Krawczynski1es1959-2004} we
determine the time fraction at which the 2-10 keV flux exceeds
50\% of the average flux (``duty cycle'', DC). In addition we
require this deviation to be significant ($S>5\sigma$). Note the
outstanding DC of Mkn\,421. Supporting the claim that variability
is a defining property of BL~Lacs, a flat distribution of the DC
in $\nu_\gamma L_\gamma$ is found (Fig.~\ref{fig:comp:compaspecstime}). A
previous study \cite{Krawczynski1es1959-2004} including Mkn~421,
Mkn~501, 1ES~2344, 1H~1426 and 1ES~1959 only had found indications
for an anticorrelation of DC and $M_\bullet$, which in our
enlarged sample is weakened mainly by the recently discovered
sources 1ES~2356, PKS~0548, and BL Lac.

Turning to the minimum VHE variability timescales $\tau$, these do
not scale with $M_\bullet$. This implies that flares originate
from a much smaller region than the BH radius and (more
importantly) that the BH properties do not influence the emission
process too much, but the jet environment may be more important.
Note that, in spite of the expected scaling behavior the TeV
blazars hosting the more massive BH, Mkn~501 and Mkn~421, seem to
exhibit the smallest $\tau$. This, however, may be a selection
effect caused (1) by their proximity, and (2) by instrumental
sensitivity, as small $\tau$ measurements require strong sources.
The latter also disables strong claims about $\tau$-luminosity
correlations yet, and all $\tau$ values are to be understood as
upper limits.

\section{Conclusions and Outlook}

The observation of VHE blazars has started to become less biased:
Not only blazars with hard spectra or in a flaring state are now
detected, but a much higher dynamical range of VHE $\gamma$
emission levels and states is probed, flare statistics studies
(e.g.~\cite{Florian}) are within reach, and generic blazar
properties start to become accessible.
Thus the era of VHE blazar astronomy has been entered---astronomy
being understood as the study of generic properties of a given
class of objects.

\section{Acknowledgments}
\vspace{-0.2cm}
The author thanks E. Lorenz, H. Meyer, and W. Bednarek for discussions and useful comments.

\end{document}